\documentclass[%
 reprint,prd,
nofootinbib,
 amsmath,amssymb,
 aps,
floatfix,
]{revtex4-2}

\usepackage{graphicx}
\usepackage{dcolumn}
\usepackage{bm}
\usepackage{xcolor}
\usepackage{ulem}
\usepackage[caption=false]{subfig}
\usepackage{lipsum}
\usepackage{tensor}
\usepackage[T1]{fontenc}

\usepackage[hidelinks]{hyperref}

\definecolor{TurkishBlue}{HTML}{144893}
\hypersetup{colorlinks=true,citecolor=TurkishBlue,linkcolor=TurkishBlue,urlcolor=TurkishBlue}

\usepackage [english]{babel}

\def\VEV#1{{\left\langle #1 \right \rangle}}

\newcommand{\hOmega}{{\hat{\Omega}}}
\DeclareSymbolFont{toneitalic}{T1}{\familydefault}{m}{it}
\DeclareMathSymbol{\crpartial}{\mathord}{toneitalic}{"F0}
\newcommand{\apjl}{{Astrophys.\ J.\ Lett.}}

\begin{document}

\preprint{}

\title{Efficient computation of overlap reduction functions for pulsar timing arrays}

\author{Neha Anil Kumar}
\email{nanilku1@jhu.edu}
\affiliation{William H.\ Miller III Department of Physics and Astronomy, Johns Hopkins University, 3400 N.\ Charles St.,
Baltimore, Maryland 21218, USA}%
\author{Marc Kamionkowski}
\email{kamion@jhu.edu}
\affiliation{William H.\ Miller III Department of Physics and Astronomy, Johns Hopkins University, 3400 N.\ Charles St.,
Baltimore, Maryland 21218, USA}%

%



\begin{abstract}
Pulsar timing arrays seek and study gravitational waves (GWs) through the angular two-point correlation function of timing residuals they induce in pulsars.  The two-point correlation function induced by the standard transverse-traceless GWs is the famous Hellings-Downs curve, a function only of the angle between the two pulsars.  Additional polarization modes (vector/scalar) that may arise in alternative-gravity theories have different angular correlation functions. Furthermore,  anisotropy, linear, or circular polarization in the stochastic GW background gives rise to additional structure in the two-point correlation function that cannot be written simply in terms of the angular separation of the two pulsars.  In this paper, we provide a simple formula for the most general two-point correlation function--or overlap reduction function (ORF)--for a gravitational-wave background with an arbitrary polarization state, possibly containing anisotropies in its intensity and polarization (linear or circular).  We provide specific expressions for the ORFs sourced by the general-relativistic transverse-traceless GW modes as well as vector (or spin-1) modes that may arise in alternative-gravity theories.
\end{abstract}

\maketitle

Recent evidence reported by several pulsar timing arrays (PTAs) for detection of the telltale Hellings-Downs (HD) \cite{1983ApJ...265L..39H} correlation expected for a stochastic gravitational-wave background (SGWB) \cite{NanoGrav:2023gor,EPTA:2023fyk,Reardon:2023gzh,Xu:2023wog} makes this a particularly exciting time to think about what more can be done with PTAs in the future.  The $\sim$nHz GWs detected by PTAs are expected from inspirals of merging supermassive black hole binaries (SMBHBs) \cite{Rajagopal:1994zj,Jaffe:2002rt,Sesana:2008mz,Phinney:2001di,Middleton:2015oda,Sato-Polito:2023spo,Gardiner:2023zzr}, but there is no shortage of ideas for GWs at these frequencies from a variety of other possible sources \cite{Olmez:2010bi, Sousa:2013aaa, Miyamoto:2012ck, Kuroyanagi:2012jf, Caprini:2010xv, Starobinsky:1979ty, Zhao:2013bba, Inomata:2016rbd, NANOGrav:2023hvm}.  It is also possible to seek the signatures of other polarization states of GWs--scalar or vector--in addition to the usual general-relativistic tensor \cite{LeeJenetPrice:2008,Chamberlin:2011ev,Gair:2015hra,Cornish:2017oic,Qin:2018yhy, Tasinato:2023zcg, Bernardo:2023jhs, Bernardo:2023pwt}. 

The  detected HD correlation, characterizing an isotropic, spin-2 background, depends only on the angular separation between two pulsars. Furthermore, scalar or vector GWs can give rise to different functional dependences on this angular separation.  If the SGWB is anisotropic \cite{Mingarelli:2013dsa,Gair:2014rwa,Taylor:2013esa,NANOGrav:2023tcn}, linearly \cite{Chu:2021krj,Liu:2022skj,AnilKumar:2023kvt} or circularly polarized \cite{Kato:2015bye,Sato-Polito:2021efu}, then there is additional structure in the two-point correlation function--or overlap reduction function (ORF), as it is known in the PTA community--that most generally depends on the sky locations of the pulsars.  The calculation of these ORFs is the subject of many papers.  While the expression for the HD curve is fairly compact, ORFs in other cases require evaluation of complicated integrals  (especially in the case of anisotropy or polarization) of the plane-wave ORFs over the entire sphere, often involving numerical evaluations. Although analytic expressions have been obtained for the lowest-order anisotropies (i.e., the dipole and quadrupole), even in these simplest cases the expressions are fairly cumbersome. Moreover, these victories are still pyrrhic, as those computational-frame expressions then need to be rotated with Wigner rotation matrices to the observer frame.

In this paper, we present a simple, easily understandable, and numerically evaluated formula [Eq.~(\ref{eqn:ORFs})] for the most general ORF.  In this formalism, the ORFs are specified by coefficients $F^L_{\ell\ell'}$ where $L$ is the angular-momentum quantum number associated with anisotropy (or $L=0$ for the isotropic component), and $\ell\ell'$ are associated with the dependence of the ORFs on the pulsar positions.  {\it All} prior ORFs can be parametrized in terms of some fairly simple $F^L_{\ell\ell'}$.  We also provide, for the first time, ORFs from intensity anisotropies and linear/circular polarization for a spin-1 SGWB.  With Eq.~(\ref{eqn:ORFs}), it should be trivial to generalize any existing PTA analysis pipeline that seeks intensity anisotropies to look for linear- or circular- polarization anisotropies in spin-2 and spin-1 SGWBs.
 
We take the spin-2 (spin-1) SGWB to be a realization of a Gaussian random field in which the Fourier amplitudes $h_p(f,\hOmega)$ of polarization $p \in \{+,\, \times\}$ ($p \in \{x,\, y\}$) and frequency $f$, propagating in the $\hOmega$ direction are chosen from a distribution with variances given by
\begin{equation}
    \langle \tilde{h}_p^*(f, \hat{\Omega}) \tilde{h}_{p'}(f', \hat{\Omega}')\rangle = \delta_D(f-f')\delta_D^2(\hat{\Omega},\hat{\Omega}')\mathcal{P}_{p, p'}(f, \hat{\Omega}),
    \label{eq: GW_strain_correlations}
\end{equation}
where the $\delta_D$'s are Dirac delta functions.
Here, $\mathcal{P}_{p, p'}$ is the spectral density of the background, which depends on the polarization of the SGWB and the propagation direction. This polarization tensor can be expressed as 
\begin{equation}
    \mathcal{P}_{p, p'}(f, \hat{\Omega}) = \begin{pmatrix} I(f, \hat{\Omega}) + Q(f, \hat{\Omega}) & U(f, \hat{\Omega}) - i V(f, \hat{\Omega}) \\ U(f, \hat{\Omega}) + i V(f, \hat{\Omega}) & I(f, \hat{\Omega}) - Q(f, \hat{\Omega})\end{pmatrix},
    \label{eq: polarization_corr_tensor}
\end{equation}
analogous to how this tensor is defined using Stokes parameters in standard electromagnetism \cite{Conneely:2018wis,Ali-Haimoud:2020iyz,Ali-Haimoud:2020ozu,AnilKumar:2023kvt}. In the above characterization of the tensor $ \mathcal{P}_{p, p'}(f, \hat{\Omega})$, we have assumed the following definitions of the Stokes parameters in terms of the GW strain:
\begin{equation}
\begin{aligned}
    I(f, \hat{\Omega}) &= \frac{1}{2}\langle|\tilde{h}_+|^2 + |\tilde{h}_\times|^2\rangle,\\
    Q(f, \hat{\Omega}) &= \frac{1}{2}\langle|\tilde{h}_+|^2 - |\tilde{h}_\times|^2\rangle,\\
    U(f, \hat{\Omega}) &= {\rm{Re}}\langle{\tilde{h}_+^*\tilde{h}_\times}\rangle = \frac{1}{2}\langle\tilde{h}_+^*\tilde{h}_\times + \tilde{h}_\times^*\tilde{h}_+ \rangle, \\
    V(f, \hat{\Omega}) &= {\rm{Im}}\langle{\tilde{h}_+^*\tilde{h}_\times}\rangle = \frac{1}{2i}\langle\tilde{h}_+^*\tilde{h}_\times - \tilde{h}_\times^*\tilde{h}_+ \rangle,
\label{eqn:stokes}    
\end{aligned}
\end{equation}
where $I(f,\hat{\Omega})$ is the intensity, $V(f,\hOmega)$ the circular polarization, and $Q(f,\hat{\Omega})$ and $U(f,\hat{\Omega})$ characterize the linear polarization\footnote{In Eq.~\ref{eqn:stokes} we have assumed a transverse-traceless SGWB. For our calculations on a spin-1 background the definition of the Stoke's parameters follows the same format with every instance of $h_+$ ($h_{\times}$) replaced with $h_{x}$ ($h_y$), and updated expansion coefficients $d_{LM}^X$ ($X\in \{I,\, V,\, +,\, -\}$) in Eqs.~(\ref{eqn:expansion_I}-\ref{eqn:expansion_QU}).}. 

At this point, we assume that the frequency and angular dependence of the maps are separable and that we are working with measurements of the SGWB in a single frequency bin. Subsequently, we drop the explicit $f$ dependence in the following analysis. The additional steps required to connect the residual correlations here with the data are exactly the same as presented in other analysis (see, for example Refs.~\cite{Sato-Polito:2023spo, Mingarelli:2013dsa}), depending on the assumed source and statistics of the background.

Under these assumptions, the angular dependence of the Stokes parameters can be expanded in terms of spherical-harmonic functions as follows:
\begin{eqnarray}
    I(\hat{\Omega}) &= &I_0\sum_{L = 0}^{\infty} \sum_{M = -L}^L c_{LM}^I Y_{LM}(\hat{\Omega})  ,
    \label{eqn:expansion_I}  \\
    V(\hat{\Omega}) &= &I_0\sum_{L = 0}^{\infty} \sum_{M = -L}^L c_{LM}^VY_{LM}(\hat{\Omega}) , 
    \label{eqn:expansion_V}  \\
    P_{\pm}(\hat{\Omega}) &= &I_0\sum_{L = L_{\rm min}}^\infty \sum_{M = -L}^L c_{LM}^{\pm} {_{\pm S}Y_{LM}}(\hat{\Omega}),
    \label{eqn:expansion_QU}    
\end{eqnarray}
where we have defined the spin-$S$ fields $P_{\pm}=(Q\pm iU)$ --expanded in terms of spin-weighted spherical harmonics ${}_{\pm S} Y_{LM}(\hOmega)$. For a transverse-traceless SGWB, $S = 4$ and thus the nonvanishing terms start at $L_{\rm min} = 4$, whereas $S = L_{\rm min} = 2$ for a spin-1 background. Moreover, the coefficient $I_0$ describes these maps averaged over the frequencies in a frequency band that is centered at $f$. We assume a normalization for $I_0$ such that $c_{00}^I = 1$. With the parametrizations above, any frequency dependence in the polarization relative to the intensity is absorbed in the frequency dependence of $c_{LM}^\pm$ or $c_{LM}^V$.


Since $Q$ and $U$ are not coordinate invariants, we rewrite the linear-polarization maps in terms of scalar $E$ and pseudoscalar $B$ modes \cite{Conneely:2018wis}, with spherical-harmonic coefficients:
\begin{align}
    c_{LM}^E = \frac{1}{2}\left(c_{LM}^+ + c_{LM}^-\right),  && c_{LM}^B = -\frac{i}{2}\left(c_{LM}^+ - c_{LM}^-\right).
    \label{eq: plus_minus_to_EB}
\end{align}
Since the scalar and pseudoscalar $E(\hat n)$ and $B(\hat n)$ fields are real, all the expansion coefficients now satisfy $c_{LM}^X  = (-1)^M c_{L,-M}^{X,*}$ for $X \in \{I,\, V,\, E,\, B\}$.

The fractional change in the arrival time of a pulse from a distant pulsar, induced by a metric perturbation $h_{ij}(t, \hat{\Omega}, {\bf{x}})$ from a GW propagating in the $\hat{\Omega}$ direction (assuming a metric theory of gravity) is given by:
\begin{equation}
    z_{a}(t|\hat{\Omega}) = \frac{n_a^i n_a^j}{2(1 + \hat{\Omega} \cdot \hat{n}_a)} \Delta h_{ij},
    \label{eq: timing_resid_oneGW}
\end{equation}
where the subscript $a$ labels a given pulsar, and $\hat n_a$ is the location of that pulsar in the sky. Here we have defined $\Delta h_{ij} \equiv h_{ij}(t_e, \hat{\Omega}, \vec{0}) - h_{ij}(t_p, \hat{\Omega}, L_a\hat{n}_a)$ as the difference between the metric perturbation arriving at the Solar System barycenter (located at $\vec{0}$) at time $t_e$, and at the pulsar (located at $L_a\hat{n}_a$) at time $t_p = t - L_a$. However, consistent with existing analysis, we disregard the pulsar term in the following calculations.

The time-sequence data for each pulsar $a$ can be Fourier transformed to provide Fourier-domain timing residuals $z_f(\hat n_a)$ (which are most generally complex) at frequency $f$. Then, the SGWB signal can be extracted from PTA data by correlating the timing residuals across pulsar pairs. In configuration space, this angular correlation takes the form,
\begin{equation}
    \langle z_f(\hat{n}_a)^*z_f(\hat{n}_b)\rangle = \sum_{X} i^X \left[I_0 \sum_{L,M}c_{LM}^X\Gamma^X_{LM}(\hat{n}_a, \hat{n}_b)\right],
    \label{eq: exp_config}
\end{equation}
where $X \in \{I,\ V\ , E,\ B\}$ and $i^X = 1$ ($i^X = i$) when $X \in \{I, E\}$ ($X \in \{V, B\}$). The functions $\Gamma^X_{LM}(\hat{n}_a, \hat{n}_b)$ are the ORFs that quantify the dependence of the signal on the locations of the two pulsars $a$ and $b$ in the sky. 

The same angular two-point correlation function can (most generally) be expanded as,
\begin{eqnarray}
    \VEV{z_f(\hat{n}_a)^* z_f(\hat{n}_b)}  &= &\sum_{\ell}\frac{2\ell + 1}{4\pi}C_\ell P_\ell(\hat{n}_a\cdot\hat{n}_b)  \nonumber \\
    &+ & \sum_{L\geq1}\sum_{M=-L}^L \sum_{\ell\ell'} A^{LM}_{\ell\ell'}\left\{Y_{\ell}(\hat{n}_a)\otimes Y_{\ell'}(\hat{n}_b)\right\}_{LM}, \nonumber \\
    \label{eq: biposh_exp_real}
\end{eqnarray}
where 
\begin{multline}
    \left\{Y_{\ell}(\hat{n}_a)\otimes Y_{\ell'}(\hat{n}_b)\right\}_{LM}  = \\    \sum_{mm'} \VEV{\ell m\ell' m'| LM}Y_{\ell m}(\hat{n}_a)Y_{\ell' m'}(\hat{n}_b),
\label{eqn:biposh}
\end{multline}
are bipolar spherical harmonics (BiPoSHs) \cite{Hajian:2003qq,Hajian:2005jh,Joshi:2009mj,Book:2011na} and $\VEV{\ell m\ell' m'| LM}$ are Clebsch-Gordan coefficients. These functions constitute a complete, orthonormal basis for functions of  $\hat{n}_a$ and $\hat{n}_b$, in terms of total-angular-momentum states of quantum numbers $LM$. Although the coefficients $A_{\ell\ell'}^{LM}$ are (anti)symmetric for $\ell + \ell' + L =$ even (odd) when the map $z(\hat{n})$ is real \cite{Book:2011na}, these symmetry properties do not hold if the timing-residual map is complex, i.e., if the SGWB has a non-zero circular polarization \cite{Belgacem:2020nda}. Note that the sum in the second term in Eq.~(\ref{eq: biposh_exp_real}) includes values of $\ell\ell'$ that satisfy the triangle inequality, $|\ell-\ell'| \leq L \leq \ell+\ell'$.

The first term in Eq.~(\ref{eq: biposh_exp_real}) represents the contribution to the correlation coming from the isotropic, rotationally invariant component of the background. The harmonic-space coefficients (power spectrum) corresponding to the HD curve are $C_\ell \propto (\ell-2)!/(\ell+2)!$ \cite{Gair:2014rwa,Roebber:2016jzl,Qin:2018yhy}. The BiPoSH coefficients $A^{LM}_{\ell\ell'}$ quantify departures from statistical isotropy \cite{Hotinli:2019tpc,Ali-Haimoud:2020iyz,Ali-Haimoud:2020ozu}, circular-polarization anisotropies \cite{Belgacem:2020nda}, and linear polarization \cite{Chu:2021krj,Liu:2022skj,AnilKumar:2023kvt}.  Each term, of given $LM$, in Eqs.~(\ref{eqn:expansion_I})--(\ref{eqn:expansion_QU}) gives rise to nonzero $A^{LM}_{\ell\ell'}$ of the same $LM$.  Intensity anisotropies and $E$-mode linear polarization are scalars and thus have non vanishing BiPoSH coefficients only for $\ell+\ell'+L=$even, while $B$-mode linear polarization and circular polarization are pseudoscalars and thus are nonvanishing only for $\ell+\ell'+L=$odd.

\begin{table*}
\centering
    \everymath{\displaystyle}
    \renewcommand{\arraystretch}{1.2}
    \resizebox{0.8\textwidth}{!}{\begin{tabular}{|c|c|c|}
    \hline
    $X$ & Spin-2 & Spin-1  \\
	\hline\hline

    $I$ & $2 z_\ell^t z_{\ell'}^t  \left( \begin{array}{ccc} \ell & \ell'& L \\ -2 & 2 &  0 \end{array} \right) X^L_{\ell\ell'}$ &  $ -2 z_\ell^v z_{\ell'}^v  \left( \begin{array}{ccc} \ell & \ell'& L \\ -1 & 1 &  0 \end{array} \right) X^L_{\ell\ell'}$  \\ \hline
 
    $V$ & $-2 z_\ell^t z_{\ell'}^t  \left( \begin{array}{ccc} \ell & \ell'& L \\ -2 & 2 &  0 \end{array} \right) (1-X^L_{\ell\ell'})$ &  $ 2z_\ell^v z_{\ell'}^v  \left( \begin{array}{ccc} \ell & \ell'& L \\ -1 & 1 &  0 \end{array} \right) (1 - X^L_{\ell\ell'})$  \\ \hline
    
    $E$ & $2 z_\ell^t z_{\ell'}^t  \left( \begin{array}{ccc} \ell & \ell'& L \\ 2 & 2 &  -4 \end{array} \right) X^L_{\ell\ell'}$ &  $ 2 z_\ell^v z_{\ell'}^v  \left( \begin{array}{ccc} \ell & \ell'& L \\ 1 & 1 &  -2 \end{array} \right) X^L_{\ell\ell'}$  \\ \hline

    $B$ & $2 z_\ell^t z_{\ell'}^t  \left( \begin{array}{ccc} \ell & \ell'& L \\ 2 & 2 &  -4 \end{array} \right) (1-X^L_{\ell\ell'})$  &  $ 2 z_\ell^v z_{\ell'}^v  \left( \begin{array}{ccc} \ell & \ell'& L \\ 1 & 1 &  -2 \end{array} \right) (1 - X^L_{\ell\ell'})$  \\ \hline
\end{tabular}}
\caption{The values of $F^{L,X}_{\ell\ell'}$ that appear in Eq.~(\ref{eqn:ORFs}) for the ORFs for intensity ($I$), circular-polarization ($V$), and $E$- and $B$-mode linear-polarization anisotropies of multipole $L$.  Here, $X_L^{\ell\ell'} = 1$ for $\ell+\ell'+L$=even and 0 otherwise. }
\label{tab:F}
\end{table*}

 Comparing Eq.~(\ref{eq: exp_config}) to Eq.~(\ref{eq: biposh_exp_real}) reveals an expression for the ORFs $\Gamma^X_{LM}(\hat{n}_a, \hat{n}_b)$ as an expansion in the BiPoSH basis:
\begin{equation}
    \Gamma_{L M}^X(\hat n_a,\hat n_b) = (-1)^L \sqrt{\pi} \sum_{\ell\ell'} F^{L,X}_{\ell\ell'} \left\{Y_{\ell}(\hat{n}_a)\otimes Y_{\ell'}(\hat{n}_b)\right\}_{LM}.
\label{eqn:ORFs} 
\end{equation}
Here, the sums start at $\ell_{\rm min} = 2$ ($\ell_{\rm min} = 1$) for spin-2 (spin-1) GWs, running over all values of $\ell$ and $\ell'$ that satisfy the triangle inequality. Although the sums over $\ell\ell'$ are nominally infinite, they can be cut off at a maximum value $\ell_{\rm max}\sim \sqrt{N_p}$.  Here, $N_p$ is the number of pulsars, and angular structures described by multipoles with larger $\ell$ cannot be resolved by the survey. All non-trivial information about the ORFs is included in the expansion coefficients $F^{L,X}_{\ell\ell'}$, which represent the fractional contribution from each $c_{LM}^X$ to the total BiPoSH amplitude $A_{\ell\ell'}^{LM}$ \cite{AnilKumar:2023kvt}. 

In several recent papers, the BiPoSH coefficients have been expressed in terms of the SGWB polarization expansion coefficients $c_{LM}^X$ \cite{Hotinli:2019tpc, Belgacem:2020nda, Liu:2022skj, Chu:2021krj, AnilKumar:2023kvt} for spin-2 gravitational waves. The results are collected in the `Spin-2' column in Table~\ref{tab:F}, where
\begin{equation}
    z_\ell^t \equiv (-1)^\ell \sqrt{\frac{4\pi (2\ell+1) (\ell-2)!}{(\ell+2)!}}.
\end{equation}
Equation~(\ref{eqn:ORFs}), along with the results in the middle column of the table, allows all prior spin-2 anisotropy/polarization ORFs to be reproduced with one simple code.

As a simple proof of concept, we point out that the rotationally invariant part of the correlation function (i.e., the HD curve) can be obtained by taking $L=0$ in Eq.~(\ref{eqn:ORFs}).  The sums over $mm'$ in the expression for the BiPoSHs [Eq.~(\ref{eqn:biposh})] yields, in this case, Legendre polynomials.  This then leads to the first term in Eq.~(\ref{eq: biposh_exp_real}).

The simplicity and elegance of these results allow us to guess the analogous ORFs for a SGWB composed of spin-1 GWs (as may arise in alternative-gravity theories).  One may guess that the $(\ell-2)!/(\ell+2)!$ in the $z_\ell^t$ would be replaced by $(\ell-1)!/(\ell+1)!$ and the 2's and 4's (which correspond, respectively, to the spin of the GW and the spin of the GW polarization) in the Wigner-$3j$ symbols be replaced by 1's and 2's, respectively. Although this gives something nearly correct, a few more minor changes are necessary.

To get the right answers, shown in the last column of Table~\ref{tab:F}, we review some elements of the calculations for spin-2 GWs.  Consider a plane wave propagating in the $\hat z$ direction with $+$ polarization.  It has a polarization tensor with nonzero components $\epsilon_{xx} = -\epsilon_{yy}=1$, and so [based on Eq.~(\ref{eq: timing_resid_oneGW})] the angular dependence of the timing residual for a pulsar (with unit GW amplitude) in the $\hat n$ direction is given by
\begin{equation}
    z_f(\hat n) = \frac{ n^i n^j}{2(1+\hat z \cdot \hat n)} \epsilon_{ij} = \frac{1}{2}(1-\cos\theta)\cos2\phi.
\end{equation}
Then, projections of this angular pattern onto spherical harmonics are \cite{Hotinli:2019tpc}
\begin{equation}
    z_{\ell m} \propto z_\ell^t (\delta_{m2}+\delta_{m,-2}).
\end{equation}
Given these coefficients, the projection of a GW propagating in any other direction can then be obtained with the appropriate Wigner rotation matrices \cite{Hotinli:2019tpc}.  

\begin{figure*}
 \centering
 \includegraphics[width=\columnwidth]{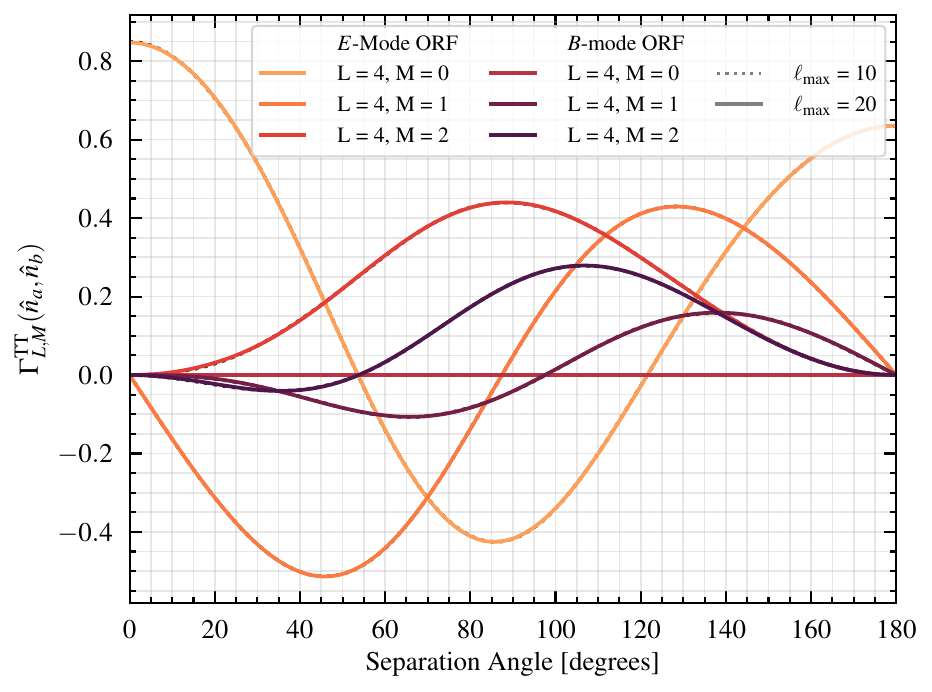}%
 \centering
 \includegraphics[width=\columnwidth]{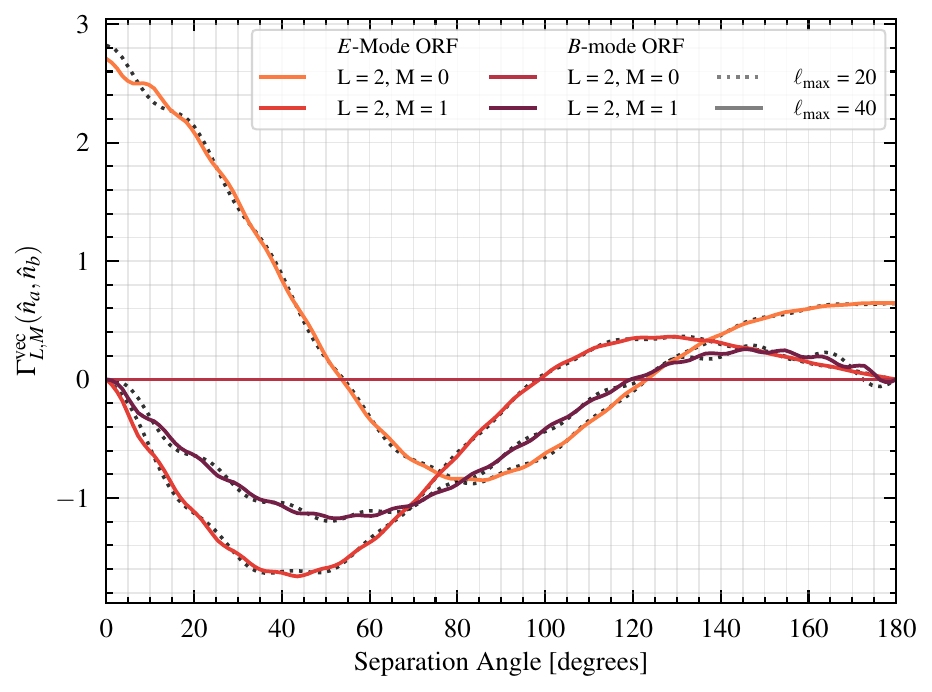}%
  \caption{\textit{Left:} example ORFs for linear-polarization ($E$- and $B$-mode) anisotropies with $L = 4$, assuming a transverse-traceless SGWB. The grey dotted lines correspond to results computed using $\ell_{\rm max} = 10$, whereas the colored solid lines correspond to $\ell_{\rm max} = 20$. The rapid convergence of the series in Eq.~\eqref{eqn:ORFs} for a spin-2 SGWB makes the dotted curve minimally visible behind the results computed with the higher $\ell_{\rm max}$ that produce (almost) identical results. \textit{Right:} example ORFs for linear-polarization ($E$- and $B$-mode) anisotropies with $L = 2$, assuming a spin-1 SGWB. The gray, dotted lines correspond to $\ell_{\rm max} = 20$, whereas the solid colored lines represent results computed to $\ell_{\rm max} = 40$. The results are computed in the computational frame, where we assume $\hat{n}_a = (0,\,0,\,1)$ such that the ORFs can be expressed simply in terms of the angle between the two pulsars $a$ and $b$. }
 \label{fig:Gamma_EB}%
\end{figure*}

The BiPoSH coefficients for any given $c^I_{LM}$ are then obtained by integrating the contributions of GWs in all directions, weighted by the angular pattern implied by the given $LM$ mode.  This involves an integration over the $LM$ spherical harmonic [as suggested by Eq.~(\ref{eqn:expansion_I})], and two rotation matrices which, for a spin-2 gravitational wave, are spin-2 spin-weighted spherical harmonics.  That integral then yields the Wigner-$3j$ symbol in the $I$ $\times$ ``Spin-2'' entry of Table~\ref{tab:F} \cite{Hotinli:2019tpc}.  The BiPoSHs are nonzero only for even $\ell+\ell'+L$, which is accounted for in our results by the factor of $X_{\ell\ell'}^L$.  
The calculation for the circular polarization is similar \cite{Belgacem:2020nda}, but uses right- and left-handed circularly polarized GWs. The result is analogous except that the BiPoSHs are nonzero for $\ell+\ell'+L=$odd, given that $V(\hat n)$ is a pseudoscalar.  Likewise, similar steps lead to the spin-2 results for linear polarization, with the appropriate Wigner-$3j$'s, this time accounting for the spin-4 nature of the linear-polarization field, in contrast to the spin-0 intensity and circular-polarization fields \cite{AnilKumar:2023kvt}. The left plot of Fig.~\ref{fig:Gamma_EB} displays a few example ORFs for the lowest-order $E$- and $B$-mode anisotropies ($L = 4$) obtained using Eq.~\eqref{eqn:ORFs}, plugging in the appropriate coefficients $F_{\ell\ell'}^L$ from the `Spin-2' column of Table~\ref{tab:F}.

Now consider a vectorial (spin-1) GW propagating in the $\hat{z}$ direction with $x$ polarization.  The polarization tensor now has nonzero components, $\epsilon_{xz}=\epsilon_{zx}=1$.  Using Eq.~(\ref{eq: timing_resid_oneGW}), the timing residual for such a GW (with unit GW amplitude) has angular dependence\footnote{The singularity in $z$ in the limit that the GW comes from {\it precisely} the direction of a pulsar leads to a divergence in the zero-lag timing-residual correlation function. However, the expressions here are still finite and good approximations of the correlation function for two different pulsars with realistic pulsar separations.  The divergence in the zero-lag correlation function is tempered by the inclusion of the pulsar term in the calculation (see, e.g., Refs. \cite{LeeJenetPrice:2008, Chamberlin:2011ev, Gair:2015hra, Cornish:2017oic,Qin:2018yhy,NANOGrav:2021ini}), and may also be tempered by subluminal GWs \cite{Bernardo:2022rif}.},
\begin{equation}
    z_f(\hat n) = \frac{\sin2\theta \cos\phi}{2(1 + \cos\theta)},
\end{equation}
which has expansion coefficients,
\begin{equation}
    z_{\ell m} \propto z_\ell^v (\delta_{m,-1} - \delta_{m,1}),
\end{equation}
where
\begin{equation}
    z_{\ell}^v \equiv (-1)^\ell \sqrt{4\pi(2\ell+1)} \left( \frac{1}{\sqrt{\ell(\ell+1)}} - \frac{\sqrt{2}}{3} \delta_{\ell 1} \right).
\label{eqn:correctvectorzl}
\end{equation}

The remaining contributions to $F^{L,X}_{\ell\ell'}$ for spin-1 are obtained by following the remaining steps in Sec. V.B of Ref.~\cite{Hotinli:2019tpc}, but replacing the spin-2 Wigner rotation matrices with spin-1 matrices.  The algebra then leads to the results in the ``Spin-1'' column of Table \ref{tab:F}.  To clarify, these are the ORFs for a spin-1 SGWB with expansions as in Eqs.~(\ref{eqn:expansion_I})--(\ref{eqn:expansion_QU}) with the expansion coefficients for the spin-1 background taken to be $d_{LM}^I$ for intensity, $d_{LM}^V$ for circular polarization, and $d_{LM}^{E,B}$ for the $E$- and $B$-mode linear polarizations (keeping in mind that the linear polarization is now a spin-2 field on the celestial sphere). The right plot of Fig.~\ref{fig:Gamma_EB} displays a few example ORFs for the lowest-order $E$- and $B$-mode anisotropies ($L = 2$) obtained using Eq.~\eqref{eqn:ORFs}, plugging in the appropriate coefficients $F_{\ell\ell'}^L$ from the `Spin-1' column of Table~\ref{tab:F}. Note that the new $\ell$-dependence of the coefficient $z_{\ell}^v$ will slow down the convergence of the sum in Eq.~(\ref{eqn:ORFs}) to the true ORFs, relative to its spin-2 counterpart. This can be seen in Fig.~\ref{fig:Gamma_EB} as the visible difference between the dotted results computed using an $\ell_{\rm max} = 20$ and the solid, colored results computed for an $\ell_{\rm max} = 40$.

The situation for scalar gravitational waves is not quite as simple.  The most general scalar GW can be decomposed into a scalar transverse (ST) mode, in which there is isotropic stretching in the direction transverse to the direction of propagation of the GW, and a scalar longitudinal (SL) mode, wherein the stretching is along the propagation direction.  For the ST modes, the timing-residual pattern $z(\hat n)$ consists of only a monopole and a dipole \cite{Qin:2018yhy}, and so there is limited information that can be obtained about a background beyond detection of an amplitude.  For the SL mode, there is a dependence on the pulsar term \cite{Cornish:2017oic,Qin:2018yhy} that gives rise to a more complicated angular pattern of timing residuals, and the discussion above for vector and tensor modes requires a bit more effort to generalize.

Overlap reduction functions are the principal theoretical result connecting models of the SGWB with PTA observables.  The simplest example is the HD curve that arises from transverse-traceless GWs, but there are additional ORFs that have been calculated to account for intensity anisotropy, circular polarization, and linear polarization, and also for isotropic SGWBs of spin-0 and spin-1 GW modes that may arise in alternative-gravity theories. In most cases, these complicated functions are computed in a fixed coordinate system, to simplify the integral calculations. Not only are these integrals difficult to compute, but this methodology often requires the additional step of rotating the obtained result into the more commonly applicable celestial frame.

In this work, we have derived a simple expression, Eq.~(\ref{eqn:ORFs}), that summarizes all the ORFs derived in previous work and also provides the first ORFs for anisotropic or polarized backgrounds of vectorial SGWBs. These expressions allow for computation in any chosen frame of reference, and despite the harmonic-space methodology the results critically do not hinge on any assumptions about the distribution of pulsars across the sky. The expression for these ORFs is easily coded and evaluated numerically. In fact, the rapid decrease of $z_\ell$ with $\ell$ (especially for spin-2; less so for spin-1) makes the series fairly rapidly convergent and evaluated well even with fairly small $\ell_{\rm max}$.  The simplicity of the expression decreases the possibility for coding errors relative to analytic expressions for the anisotropy or polarization ORFs that have been provided earlier.  The uniformity of the result for all of the different possible ORFs also reduces the possibility of errors that may arise from inconsistencies in notation or conventions.  The structure of Eq.~(\ref{eqn:ORFs}) also makes clear the most general possible form an ORF can take and thus provides a simple way to map model predictions to observables: One simply specifies $F^L_{\ell\ell'}$.

The same machinery can be generalized to apply to the two-point correlation function for astrometric deflections from GWs \cite{Book:2010pf}.  These deflections can be decomposed into $E$- and $B$-modes ({\it not} to be confused with the $E$/$B$ decomposition of the SGWB linear polarization).  As a result, there will be a total of six astrometry $\times$ PTA two-point correlation functions, and thus six (or just four if parity is conserved) ORFs for each $c_{LM}^{I,V,E,B}$. It should also be possible to generalize this formalism to subluminal GWs; in this case, Eq. (\ref{eqn:ORFs}) will be preserved, but the values of the coefficients $F^L_{\ell\ell’}$ will be changed. We leave the complete enumeration of these to future work.

\smallskip
\textit{Acknowledgements}--We thank T.\ Smith, K. Inomata, G.\ Sato-Polito, M. \c{C}al{\i}\c{s}kan, and L.\ Ji for their useful comments.  This work was supported by NSF Grant No.\ 2112699 and the Simons Foundation.

\bibliography{PrettyORFs.bib}

\providecommand{\noopsort}[1]{}\providecommand{\singleletter}[1]{#1}%
\begin{thebibliography}{51}%
\makeatletter
\providecommand \@ifxundefined [1]{%
 \@ifx{#1\undefined}
}%
\providecommand \@ifnum [1]{%
 \ifnum #1\expandafter \@firstoftwo
 \else \expandafter \@secondoftwo
 \fi
}%
\providecommand \@ifx [1]{%
 \ifx #1\expandafter \@firstoftwo
 \else \expandafter \@secondoftwo
 \fi
}%
\providecommand \natexlab [1]{#1}%
\providecommand \enquote  [1]{``#1''}%
\providecommand \bibnamefont  [1]{#1}%
\providecommand \bibfnamefont [1]{#1}%
\providecommand \citenamefont [1]{#1}%
\providecommand \href@noop [0]{\@secondoftwo}%
\providecommand \href [0]{\begingroup \@sanitize@url \@href}%
\providecommand \@href[1]{\@@startlink{#1}\@@href}%
\providecommand \@@href[1]{\endgroup#1\@@endlink}%
\providecommand \@sanitize@url [0]{\catcode `\\12\catcode `\$12\catcode `\&12\catcode `\#12\catcode `\^12\catcode `\_12\catcode `\%12\relax}%
\providecommand \@@startlink[1]{}%
\providecommand \@@endlink[0]{}%
\providecommand \url  [0]{\begingroup\@sanitize@url \@url }%
\providecommand \@url [1]{\endgroup\@href {#1}{\urlprefix }}%
\providecommand \urlprefix  [0]{URL }%
\providecommand \Eprint [0]{\href }%
\providecommand \doibase [0]{https://doi.org/}%
\providecommand \selectlanguage [0]{\@gobble}%
\providecommand \bibinfo  [0]{\@secondoftwo}%
\providecommand \bibfield  [0]{\@secondoftwo}%
\providecommand \translation [1]{[#1]}%
\providecommand \BibitemOpen [0]{}%
\providecommand \bibitemStop [0]{}%
\providecommand \bibitemNoStop [0]{.\EOS\space}%
\providecommand \EOS [0]{\spacefactor3000\relax}%
\providecommand \BibitemShut  [1]{\csname bibitem#1\endcsname}%
\let\auto@bib@innerbib\@empty
\bibitem [{\citenamefont {{Hellings}}\ and\ \citenamefont {{Downs}}(1983)}]{1983ApJ...265L..39H}%
  \BibitemOpen
  \bibfield  {author} {\bibinfo {author} {\bibfnamefont {R.~W.}\ \bibnamefont {{Hellings}}}\ and\ \bibinfo {author} {\bibfnamefont {G.~S.}\ \bibnamefont {{Downs}}},\ }\bibfield  {title} {\bibinfo {title} {{Upper limits on the isotropic gravitational radiation background from pulsar timing analysis.}},\ }\href {https://doi.org/10.1086/183954} {\bibfield  {journal} {\bibinfo  {journal} {\apjl}\ }\textbf {\bibinfo {volume} {265}},\ \bibinfo {pages} {L39} (\bibinfo {year} {1983})}\BibitemShut {NoStop}%
\bibitem [{\citenamefont {Agazie}\ \emph {et~al.}(2023{\natexlab{a}})\citenamefont {Agazie} \emph {et~al.}}]{NanoGrav:2023gor}%
  \BibitemOpen
  \bibfield  {author} {\bibinfo {author} {\bibfnamefont {G.}~\bibnamefont {Agazie}} \emph {et~al.} (\bibinfo {collaboration} {NANOGrav}),\ }\bibfield  {title} {\bibinfo {title} {{The NANOGrav 15 yr Data Set: Evidence for a Gravitational-wave Background}},\ }\href {https://doi.org/10.3847/2041-8213/acdac6} {\bibfield  {journal} {\bibinfo  {journal} {Astrophys. J. Lett.}\ }\textbf {\bibinfo {volume} {951}},\ \bibinfo {pages} {L8} (\bibinfo {year} {2023}{\natexlab{a}})},\ \Eprint {https://arxiv.org/abs/2306.16213} {arXiv:2306.16213 [astro-ph.HE]} \BibitemShut {NoStop}%
\bibitem [{\citenamefont {Antoniadis}\ \emph {et~al.}(2023)\citenamefont {Antoniadis} \emph {et~al.}}]{EPTA:2023fyk}%
  \BibitemOpen
  \bibfield  {author} {\bibinfo {author} {\bibfnamefont {J.}~\bibnamefont {Antoniadis}} \emph {et~al.} (\bibinfo {collaboration} {EPTA}),\ }\bibfield  {title} {\bibinfo {title} {{The second data release from the European Pulsar Timing Array III. Search for gravitational wave signals}},\ }\href {https://doi.org/10.1051/0004-6361/202346844} {\bibfield  {journal} {\bibinfo  {journal} {Astron. Astrophys.}\ }\textbf {\bibinfo {volume} {678}},\ \bibinfo {pages} {A50} (\bibinfo {year} {2023})},\ \Eprint {https://arxiv.org/abs/2306.16214} {arXiv:2306.16214 [astro-ph.HE]} \BibitemShut {NoStop}%
\bibitem [{\citenamefont {Reardon}\ \emph {et~al.}(2023)\citenamefont {Reardon} \emph {et~al.}}]{Reardon:2023gzh}%
  \BibitemOpen
  \bibfield  {author} {\bibinfo {author} {\bibfnamefont {D.~J.}\ \bibnamefont {Reardon}} \emph {et~al.},\ }\bibfield  {title} {\bibinfo {title} {{Search for an Isotropic Gravitational-wave Background with the Parkes Pulsar Timing Array}},\ }\href {https://doi.org/10.3847/2041-8213/acdd02} {\bibfield  {journal} {\bibinfo  {journal} {Astrophys. J. Lett.}\ }\textbf {\bibinfo {volume} {951}},\ \bibinfo {pages} {L6} (\bibinfo {year} {2023})},\ \Eprint {https://arxiv.org/abs/2306.16215} {arXiv:2306.16215 [astro-ph.HE]} \BibitemShut {NoStop}%
\bibitem [{\citenamefont {Xu}\ \emph {et~al.}(2023)\citenamefont {Xu} \emph {et~al.}}]{Xu:2023wog}%
  \BibitemOpen
  \bibfield  {author} {\bibinfo {author} {\bibfnamefont {H.}~\bibnamefont {Xu}} \emph {et~al.},\ }\bibfield  {title} {\bibinfo {title} {{Searching for the Nano-Hertz Stochastic Gravitational Wave Background with the Chinese Pulsar Timing Array Data Release I}},\ }\href {https://doi.org/10.1088/1674-4527/acdfa5} {\bibfield  {journal} {\bibinfo  {journal} {Res. Astron. Astrophys.}\ }\textbf {\bibinfo {volume} {23}},\ \bibinfo {pages} {075024} (\bibinfo {year} {2023})},\ \Eprint {https://arxiv.org/abs/2306.16216} {arXiv:2306.16216 [astro-ph.HE]} \BibitemShut {NoStop}%
\bibitem [{\citenamefont {Rajagopal}\ and\ \citenamefont {Romani}(1995)}]{Rajagopal:1994zj}%
  \BibitemOpen
  \bibfield  {author} {\bibinfo {author} {\bibfnamefont {M.}~\bibnamefont {Rajagopal}}\ and\ \bibinfo {author} {\bibfnamefont {R.~W.}\ \bibnamefont {Romani}},\ }\bibfield  {title} {\bibinfo {title} {{Ultralow frequency gravitational radiation from massive black hole binaries}},\ }\href {https://doi.org/10.1086/175813} {\bibfield  {journal} {\bibinfo  {journal} {Astrophys. J.}\ }\textbf {\bibinfo {volume} {446}},\ \bibinfo {pages} {543} (\bibinfo {year} {1995})},\ \Eprint {https://arxiv.org/abs/astro-ph/9412038} {arXiv:astro-ph/9412038} \BibitemShut {NoStop}%
\bibitem [{\citenamefont {Jaffe}\ and\ \citenamefont {Backer}(2003)}]{Jaffe:2002rt}%
  \BibitemOpen
  \bibfield  {author} {\bibinfo {author} {\bibfnamefont {A.~H.}\ \bibnamefont {Jaffe}}\ and\ \bibinfo {author} {\bibfnamefont {D.~C.}\ \bibnamefont {Backer}},\ }\bibfield  {title} {\bibinfo {title} {{Gravitational waves probe the coalescence rate of massive black hole binaries}},\ }\href {https://doi.org/10.1086/345443} {\bibfield  {journal} {\bibinfo  {journal} {Astrophys. J.}\ }\textbf {\bibinfo {volume} {583}},\ \bibinfo {pages} {616} (\bibinfo {year} {2003})},\ \Eprint {https://arxiv.org/abs/astro-ph/0210148} {arXiv:astro-ph/0210148} \BibitemShut {NoStop}%
\bibitem [{\citenamefont {Sesana}\ \emph {et~al.}(2008)\citenamefont {Sesana}, \citenamefont {Vecchio},\ and\ \citenamefont {Colacino}}]{Sesana:2008mz}%
  \BibitemOpen
  \bibfield  {author} {\bibinfo {author} {\bibfnamefont {A.}~\bibnamefont {Sesana}}, \bibinfo {author} {\bibfnamefont {A.}~\bibnamefont {Vecchio}},\ and\ \bibinfo {author} {\bibfnamefont {C.~N.}\ \bibnamefont {Colacino}},\ }\bibfield  {title} {\bibinfo {title} {{The stochastic gravitational-wave background from massive black hole binary systems: implications for observations with Pulsar Timing Arrays}},\ }\href {https://doi.org/10.1111/j.1365-2966.2008.13682.x} {\bibfield  {journal} {\bibinfo  {journal} {Mon. Not. Roy. Astron. Soc.}\ }\textbf {\bibinfo {volume} {390}},\ \bibinfo {pages} {192} (\bibinfo {year} {2008})},\ \Eprint {https://arxiv.org/abs/0804.4476} {arXiv:0804.4476 [astro-ph]} \BibitemShut {NoStop}%
\bibitem [{\citenamefont {Phinney}(2001)}]{Phinney:2001di}%
  \BibitemOpen
  \bibfield  {author} {\bibinfo {author} {\bibfnamefont {E.~S.}\ \bibnamefont {Phinney}},\ }\bibfield  {title} {\bibinfo {title} {{A Practical theorem on gravitational wave backgrounds}},\ }\href@noop {} {\  (\bibinfo {year} {2001})},\ \Eprint {https://arxiv.org/abs/astro-ph/0108028} {arXiv:astro-ph/0108028} \BibitemShut {NoStop}%
\bibitem [{\citenamefont {Middleton}\ \emph {et~al.}(2016)\citenamefont {Middleton}, \citenamefont {Del~Pozzo}, \citenamefont {Farr}, \citenamefont {Sesana},\ and\ \citenamefont {Vecchio}}]{Middleton:2015oda}%
  \BibitemOpen
  \bibfield  {author} {\bibinfo {author} {\bibfnamefont {H.}~\bibnamefont {Middleton}}, \bibinfo {author} {\bibfnamefont {W.}~\bibnamefont {Del~Pozzo}}, \bibinfo {author} {\bibfnamefont {W.~M.}\ \bibnamefont {Farr}}, \bibinfo {author} {\bibfnamefont {A.}~\bibnamefont {Sesana}},\ and\ \bibinfo {author} {\bibfnamefont {A.}~\bibnamefont {Vecchio}},\ }\bibfield  {title} {\bibinfo {title} {{Astrophysical constraints on massive black hole binary evolution from Pulsar Timing Arrays}},\ }\href {https://doi.org/10.1093/mnrasl/slv150} {\bibfield  {journal} {\bibinfo  {journal} {Mon. Not. Roy. Astron. Soc.}\ }\textbf {\bibinfo {volume} {455}},\ \bibinfo {pages} {L72} (\bibinfo {year} {2016})},\ \Eprint {https://arxiv.org/abs/1507.00992} {arXiv:1507.00992 [astro-ph.CO]} \BibitemShut {NoStop}%
\bibitem [{\citenamefont {Sato-Polito}\ and\ \citenamefont {Kamionkowski}(2023)}]{Sato-Polito:2023spo}%
  \BibitemOpen
  \bibfield  {author} {\bibinfo {author} {\bibfnamefont {G.}~\bibnamefont {Sato-Polito}}\ and\ \bibinfo {author} {\bibfnamefont {M.}~\bibnamefont {Kamionkowski}},\ }\bibfield  {title} {\bibinfo {title} {{Exploring the spectrum of stochastic gravitational-wave anisotropies with pulsar timing arrays}},\ }\href@noop {} {\  (\bibinfo {year} {2023})},\ \Eprint {https://arxiv.org/abs/2305.05690} {arXiv:2305.05690 [astro-ph.CO]} \BibitemShut {NoStop}%
\bibitem [{\citenamefont {Gardiner}\ \emph {et~al.}(2023)\citenamefont {Gardiner}, \citenamefont {Kelley}, \citenamefont {Lemke},\ and\ \citenamefont {Mitridate}}]{Gardiner:2023zzr}%
  \BibitemOpen
  \bibfield  {author} {\bibinfo {author} {\bibfnamefont {E.~C.}\ \bibnamefont {Gardiner}}, \bibinfo {author} {\bibfnamefont {L.~Z.}\ \bibnamefont {Kelley}}, \bibinfo {author} {\bibfnamefont {A.-M.}\ \bibnamefont {Lemke}},\ and\ \bibinfo {author} {\bibfnamefont {A.}~\bibnamefont {Mitridate}},\ }\bibfield  {title} {\bibinfo {title} {{Beyond the Background: Gravitational Wave Anisotropy and Continuous Waves from Supermassive Black Hole Binaries}},\ }\href@noop {} {\  (\bibinfo {year} {2023})},\ \Eprint {https://arxiv.org/abs/2309.07227} {arXiv:2309.07227 [astro-ph.HE]} \BibitemShut {NoStop}%
\bibitem [{\citenamefont {Olmez}\ \emph {et~al.}(2010)\citenamefont {Olmez}, \citenamefont {Mandic},\ and\ \citenamefont {Siemens}}]{Olmez:2010bi}%
  \BibitemOpen
  \bibfield  {author} {\bibinfo {author} {\bibfnamefont {S.}~\bibnamefont {Olmez}}, \bibinfo {author} {\bibfnamefont {V.}~\bibnamefont {Mandic}},\ and\ \bibinfo {author} {\bibfnamefont {X.}~\bibnamefont {Siemens}},\ }\bibfield  {title} {\bibinfo {title} {{Gravitational-Wave Stochastic Background from Kinks and Cusps on Cosmic Strings}},\ }\href {https://doi.org/10.1103/PhysRevD.81.104028} {\bibfield  {journal} {\bibinfo  {journal} {Phys. Rev. D}\ }\textbf {\bibinfo {volume} {81}},\ \bibinfo {pages} {104028} (\bibinfo {year} {2010})},\ \Eprint {https://arxiv.org/abs/1004.0890} {arXiv:1004.0890 [astro-ph.CO]} \BibitemShut {NoStop}%
\bibitem [{\citenamefont {Sousa}\ and\ \citenamefont {Avelino}(2013)}]{Sousa:2013aaa}%
  \BibitemOpen
  \bibfield  {author} {\bibinfo {author} {\bibfnamefont {L.}~\bibnamefont {Sousa}}\ and\ \bibinfo {author} {\bibfnamefont {P.~P.}\ \bibnamefont {Avelino}},\ }\bibfield  {title} {\bibinfo {title} {{Stochastic Gravitational Wave Background generated by Cosmic String Networks: Velocity-Dependent One-Scale model versus Scale-Invariant Evolution}},\ }\href {https://doi.org/10.1103/PhysRevD.88.023516} {\bibfield  {journal} {\bibinfo  {journal} {Phys. Rev. D}\ }\textbf {\bibinfo {volume} {88}},\ \bibinfo {pages} {023516} (\bibinfo {year} {2013})},\ \Eprint {https://arxiv.org/abs/1304.2445} {arXiv:1304.2445 [astro-ph.CO]} \BibitemShut {NoStop}%
\bibitem [{\citenamefont {Miyamoto}\ and\ \citenamefont {Nakayama}(2013)}]{Miyamoto:2012ck}%
  \BibitemOpen
  \bibfield  {author} {\bibinfo {author} {\bibfnamefont {K.}~\bibnamefont {Miyamoto}}\ and\ \bibinfo {author} {\bibfnamefont {K.}~\bibnamefont {Nakayama}},\ }\bibfield  {title} {\bibinfo {title} {{Cosmological and astrophysical constraints on superconducting cosmic strings}},\ }\href {https://doi.org/10.1088/1475-7516/2013/07/012} {\bibfield  {journal} {\bibinfo  {journal} {JCAP}\ }\textbf {\bibinfo {volume} {07}},\ \bibinfo {pages} {012}},\ \Eprint {https://arxiv.org/abs/1212.6687} {arXiv:1212.6687 [astro-ph.CO]} \BibitemShut {NoStop}%
\bibitem [{\citenamefont {Kuroyanagi}\ \emph {et~al.}(2013)\citenamefont {Kuroyanagi}, \citenamefont {Miyamoto}, \citenamefont {Sekiguchi}, \citenamefont {Takahashi},\ and\ \citenamefont {Silk}}]{Kuroyanagi:2012jf}%
  \BibitemOpen
  \bibfield  {author} {\bibinfo {author} {\bibfnamefont {S.}~\bibnamefont {Kuroyanagi}}, \bibinfo {author} {\bibfnamefont {K.}~\bibnamefont {Miyamoto}}, \bibinfo {author} {\bibfnamefont {T.}~\bibnamefont {Sekiguchi}}, \bibinfo {author} {\bibfnamefont {K.}~\bibnamefont {Takahashi}},\ and\ \bibinfo {author} {\bibfnamefont {J.}~\bibnamefont {Silk}},\ }\bibfield  {title} {\bibinfo {title} {{Forecast constraints on cosmic strings from future CMB, pulsar timing and gravitational wave direct detection experiments}},\ }\href {https://doi.org/10.1103/PhysRevD.87.023522} {\bibfield  {journal} {\bibinfo  {journal} {Phys. Rev. D}\ }\textbf {\bibinfo {volume} {87}},\ \bibinfo {pages} {023522} (\bibinfo {year} {2013})},\ \bibinfo {note} {[Erratum: Phys.Rev.D 87, 069903 (2013)]},\ \Eprint {https://arxiv.org/abs/1210.2829} {arXiv:1210.2829 [astro-ph.CO]} \BibitemShut {NoStop}%
\bibitem [{\citenamefont {Caprini}\ \emph {et~al.}(2010)\citenamefont {Caprini}, \citenamefont {Durrer},\ and\ \citenamefont {Siemens}}]{Caprini:2010xv}%
  \BibitemOpen
  \bibfield  {author} {\bibinfo {author} {\bibfnamefont {C.}~\bibnamefont {Caprini}}, \bibinfo {author} {\bibfnamefont {R.}~\bibnamefont {Durrer}},\ and\ \bibinfo {author} {\bibfnamefont {X.}~\bibnamefont {Siemens}},\ }\bibfield  {title} {\bibinfo {title} {{Detection of gravitational waves from the QCD phase transition with pulsar timing arrays}},\ }\href {https://doi.org/10.1103/PhysRevD.82.063511} {\bibfield  {journal} {\bibinfo  {journal} {Phys. Rev. D}\ }\textbf {\bibinfo {volume} {82}},\ \bibinfo {pages} {063511} (\bibinfo {year} {2010})},\ \Eprint {https://arxiv.org/abs/1007.1218} {arXiv:1007.1218 [astro-ph.CO]} \BibitemShut {NoStop}%
\bibitem [{\citenamefont {Starobinsky}(1979)}]{Starobinsky:1979ty}%
  \BibitemOpen
  \bibfield  {author} {\bibinfo {author} {\bibfnamefont {A.~A.}\ \bibnamefont {Starobinsky}},\ }\bibfield  {title} {\bibinfo {title} {{Spectrum of relict gravitational radiation and the early state of the universe}},\ }\href@noop {} {\bibfield  {journal} {\bibinfo  {journal} {JETP Lett.}\ }\textbf {\bibinfo {volume} {30}},\ \bibinfo {pages} {682} (\bibinfo {year} {1979})}\BibitemShut {NoStop}%
\bibitem [{\citenamefont {Zhao}\ \emph {et~al.}(2013)\citenamefont {Zhao}, \citenamefont {Zhang}, \citenamefont {You},\ and\ \citenamefont {Zhu}}]{Zhao:2013bba}%
  \BibitemOpen
  \bibfield  {author} {\bibinfo {author} {\bibfnamefont {W.}~\bibnamefont {Zhao}}, \bibinfo {author} {\bibfnamefont {Y.}~\bibnamefont {Zhang}}, \bibinfo {author} {\bibfnamefont {X.-P.}\ \bibnamefont {You}},\ and\ \bibinfo {author} {\bibfnamefont {Z.-H.}\ \bibnamefont {Zhu}},\ }\bibfield  {title} {\bibinfo {title} {{Constraints of relic gravitational waves by pulsar timing arrays: Forecasts for the FAST and SKA projects}},\ }\href {https://doi.org/10.1103/PhysRevD.87.124012} {\bibfield  {journal} {\bibinfo  {journal} {Phys. Rev. D}\ }\textbf {\bibinfo {volume} {87}},\ \bibinfo {pages} {124012} (\bibinfo {year} {2013})},\ \Eprint {https://arxiv.org/abs/1303.6718} {arXiv:1303.6718 [astro-ph.CO]} \BibitemShut {NoStop}%
\bibitem [{\citenamefont {Inomata}\ \emph {et~al.}(2017)\citenamefont {Inomata}, \citenamefont {Kawasaki}, \citenamefont {Mukaida}, \citenamefont {Tada},\ and\ \citenamefont {Yanagida}}]{Inomata:2016rbd}%
  \BibitemOpen
  \bibfield  {author} {\bibinfo {author} {\bibfnamefont {K.}~\bibnamefont {Inomata}}, \bibinfo {author} {\bibfnamefont {M.}~\bibnamefont {Kawasaki}}, \bibinfo {author} {\bibfnamefont {K.}~\bibnamefont {Mukaida}}, \bibinfo {author} {\bibfnamefont {Y.}~\bibnamefont {Tada}},\ and\ \bibinfo {author} {\bibfnamefont {T.~T.}\ \bibnamefont {Yanagida}},\ }\bibfield  {title} {\bibinfo {title} {{Inflationary primordial black holes for the LIGO gravitational wave events and pulsar timing array experiments}},\ }\href {https://doi.org/10.1103/PhysRevD.95.123510} {\bibfield  {journal} {\bibinfo  {journal} {Phys. Rev. D}\ }\textbf {\bibinfo {volume} {95}},\ \bibinfo {pages} {123510} (\bibinfo {year} {2017})},\ \Eprint {https://arxiv.org/abs/1611.06130} {arXiv:1611.06130 [astro-ph.CO]} \BibitemShut {NoStop}%
\bibitem [{\citenamefont {Afzal}\ \emph {et~al.}(2023)\citenamefont {Afzal} \emph {et~al.}}]{NANOGrav:2023hvm}%
  \BibitemOpen
  \bibfield  {author} {\bibinfo {author} {\bibfnamefont {A.}~\bibnamefont {Afzal}} \emph {et~al.} (\bibinfo {collaboration} {NANOGrav}),\ }\bibfield  {title} {\bibinfo {title} {{The NANOGrav 15 yr Data Set: Search for Signals from New Physics}},\ }\href {https://doi.org/10.3847/2041-8213/acdc91} {\bibfield  {journal} {\bibinfo  {journal} {Astrophys. J. Lett.}\ }\textbf {\bibinfo {volume} {951}},\ \bibinfo {pages} {L11} (\bibinfo {year} {2023})},\ \Eprint {https://arxiv.org/abs/2306.16219} {arXiv:2306.16219 [astro-ph.HE]} \BibitemShut {NoStop}%
\bibitem [{\citenamefont {Lee}\ \emph {et~al.}(2008)\citenamefont {Lee}, \citenamefont {A.},\ and\ \citenamefont {Price}}]{LeeJenetPrice:2008}%
  \BibitemOpen
  \bibfield  {author} {\bibinfo {author} {\bibfnamefont {K.~J.}\ \bibnamefont {Lee}}, \bibinfo {author} {\bibfnamefont {J.~F.}\ \bibnamefont {A.}},\ and\ \bibinfo {author} {\bibfnamefont {R.~H.}\ \bibnamefont {Price}},\ }\bibfield  {title} {\bibinfo {title} {{Pulsar timing as a probe of non-einsteinian polarizations of gravitational waves}},\ }\href@noop {} {\bibfield  {journal} {\bibinfo  {journal} {Astrophys. J.}\ }\textbf {\bibinfo {volume} {685}},\ \bibinfo {pages} {1304} (\bibinfo {year} {2008})}\BibitemShut {NoStop}%
\bibitem [{\citenamefont {Chamberlin}\ and\ \citenamefont {Siemens}(2012)}]{Chamberlin:2011ev}%
  \BibitemOpen
  \bibfield  {author} {\bibinfo {author} {\bibfnamefont {S.~J.}\ \bibnamefont {Chamberlin}}\ and\ \bibinfo {author} {\bibfnamefont {X.}~\bibnamefont {Siemens}},\ }\bibfield  {title} {\bibinfo {title} {{Stochastic backgrounds in alternative theories of gravity: overlap reduction functions for pulsar timing arrays}},\ }\href {https://doi.org/10.1103/PhysRevD.85.082001} {\bibfield  {journal} {\bibinfo  {journal} {Phys. Rev. D}\ }\textbf {\bibinfo {volume} {85}},\ \bibinfo {pages} {082001} (\bibinfo {year} {2012})},\ \Eprint {https://arxiv.org/abs/1111.5661} {arXiv:1111.5661 [astro-ph.HE]} \BibitemShut {NoStop}%
\bibitem [{\citenamefont {Gair}\ \emph {et~al.}(2015)\citenamefont {Gair}, \citenamefont {Romano},\ and\ \citenamefont {Taylor}}]{Gair:2015hra}%
  \BibitemOpen
  \bibfield  {author} {\bibinfo {author} {\bibfnamefont {J.~R.}\ \bibnamefont {Gair}}, \bibinfo {author} {\bibfnamefont {J.~D.}\ \bibnamefont {Romano}},\ and\ \bibinfo {author} {\bibfnamefont {S.~R.}\ \bibnamefont {Taylor}},\ }\bibfield  {title} {\bibinfo {title} {{Mapping gravitational-wave backgrounds of arbitrary polarisation using pulsar timing arrays}},\ }\href {https://doi.org/10.1103/PhysRevD.92.102003} {\bibfield  {journal} {\bibinfo  {journal} {Phys. Rev. D}\ }\textbf {\bibinfo {volume} {92}},\ \bibinfo {pages} {102003} (\bibinfo {year} {2015})},\ \Eprint {https://arxiv.org/abs/1506.08668} {arXiv:1506.08668 [gr-qc]} \BibitemShut {NoStop}%
\bibitem [{\citenamefont {Cornish}\ \emph {et~al.}(2018)\citenamefont {Cornish}, \citenamefont {O'Beirne}, \citenamefont {Taylor},\ and\ \citenamefont {Yunes}}]{Cornish:2017oic}%
  \BibitemOpen
  \bibfield  {author} {\bibinfo {author} {\bibfnamefont {N.~J.}\ \bibnamefont {Cornish}}, \bibinfo {author} {\bibfnamefont {L.}~\bibnamefont {O'Beirne}}, \bibinfo {author} {\bibfnamefont {S.~R.}\ \bibnamefont {Taylor}},\ and\ \bibinfo {author} {\bibfnamefont {N.}~\bibnamefont {Yunes}},\ }\bibfield  {title} {\bibinfo {title} {{Constraining alternative theories of gravity using pulsar timing arrays}},\ }\href {https://doi.org/10.1103/PhysRevLett.120.181101} {\bibfield  {journal} {\bibinfo  {journal} {Phys. Rev. Lett.}\ }\textbf {\bibinfo {volume} {120}},\ \bibinfo {pages} {181101} (\bibinfo {year} {2018})},\ \Eprint {https://arxiv.org/abs/1712.07132} {arXiv:1712.07132 [gr-qc]} \BibitemShut {NoStop}%
\bibitem [{\citenamefont {Qin}\ \emph {et~al.}(2019)\citenamefont {Qin}, \citenamefont {Boddy}, \citenamefont {Kamionkowski},\ and\ \citenamefont {Dai}}]{Qin:2018yhy}%
  \BibitemOpen
  \bibfield  {author} {\bibinfo {author} {\bibfnamefont {W.}~\bibnamefont {Qin}}, \bibinfo {author} {\bibfnamefont {K.~K.}\ \bibnamefont {Boddy}}, \bibinfo {author} {\bibfnamefont {M.}~\bibnamefont {Kamionkowski}},\ and\ \bibinfo {author} {\bibfnamefont {L.}~\bibnamefont {Dai}},\ }\bibfield  {title} {\bibinfo {title} {{Pulsar-timing arrays, astrometry, and gravitational waves}},\ }\href {https://doi.org/10.1103/PhysRevD.99.063002} {\bibfield  {journal} {\bibinfo  {journal} {Phys. Rev. D}\ }\textbf {\bibinfo {volume} {99}},\ \bibinfo {pages} {063002} (\bibinfo {year} {2019})},\ \Eprint {https://arxiv.org/abs/1810.02369} {arXiv:1810.02369 [astro-ph.CO]} \BibitemShut {NoStop}%
\bibitem [{\citenamefont {Tasinato}(2023)}]{Tasinato:2023zcg}%
  \BibitemOpen
  \bibfield  {author} {\bibinfo {author} {\bibfnamefont {G.}~\bibnamefont {Tasinato}},\ }\bibfield  {title} {\bibinfo {title} {{Kinematic anisotropies and pulsar timing arrays}},\ }\href {https://doi.org/10.1103/PhysRevD.108.103521} {\bibfield  {journal} {\bibinfo  {journal} {Phys. Rev. D}\ }\textbf {\bibinfo {volume} {108}},\ \bibinfo {pages} {103521} (\bibinfo {year} {2023})},\ \Eprint {https://arxiv.org/abs/2309.00403} {arXiv:2309.00403 [gr-qc]} \BibitemShut {NoStop}%
\bibitem [{\citenamefont {Bernardo}\ \emph {et~al.}(2023)\citenamefont {Bernardo}, \citenamefont {Liu},\ and\ \citenamefont {Ng}}]{Bernardo:2023jhs}%
  \BibitemOpen
  \bibfield  {author} {\bibinfo {author} {\bibfnamefont {R.~C.}\ \bibnamefont {Bernardo}}, \bibinfo {author} {\bibfnamefont {G.-C.}\ \bibnamefont {Liu}},\ and\ \bibinfo {author} {\bibfnamefont {K.-W.}\ \bibnamefont {Ng}},\ }\bibfield  {title} {\bibinfo {title} {{Correlations for an anisotropic polarized stochastic gravitational wave background in pulsar timing arrays}},\ }\href@noop {} {\  (\bibinfo {year} {2023})},\ \Eprint {https://arxiv.org/abs/2312.03383} {arXiv:2312.03383 [gr-qc]} \BibitemShut {NoStop}%
\bibitem [{\citenamefont {Bernardo}\ and\ \citenamefont {Ng}(2023{\natexlab{a}})}]{Bernardo:2023pwt}%
  \BibitemOpen
  \bibfield  {author} {\bibinfo {author} {\bibfnamefont {R.~C.}\ \bibnamefont {Bernardo}}\ and\ \bibinfo {author} {\bibfnamefont {K.-W.}\ \bibnamefont {Ng}},\ }\bibfield  {title} {\bibinfo {title} {{Testing gravity with cosmic variance-limited pulsar timing array correlations}},\ }\href@noop {} {\  (\bibinfo {year} {2023}{\natexlab{a}})},\ \Eprint {https://arxiv.org/abs/2306.13593} {arXiv:2306.13593 [gr-qc]} \BibitemShut {NoStop}%
\bibitem [{\citenamefont {Mingarelli}\ \emph {et~al.}(2013)\citenamefont {Mingarelli}, \citenamefont {Sidery}, \citenamefont {Mandel},\ and\ \citenamefont {Vecchio}}]{Mingarelli:2013dsa}%
  \BibitemOpen
  \bibfield  {author} {\bibinfo {author} {\bibfnamefont {C.~M.~F.}\ \bibnamefont {Mingarelli}}, \bibinfo {author} {\bibfnamefont {T.}~\bibnamefont {Sidery}}, \bibinfo {author} {\bibfnamefont {I.}~\bibnamefont {Mandel}},\ and\ \bibinfo {author} {\bibfnamefont {A.}~\bibnamefont {Vecchio}},\ }\bibfield  {title} {\bibinfo {title} {{Characterizing gravitational wave stochastic background anisotropy with pulsar timing arrays}},\ }\href {https://doi.org/10.1103/PhysRevD.88.062005} {\bibfield  {journal} {\bibinfo  {journal} {Phys. Rev. D}\ }\textbf {\bibinfo {volume} {88}},\ \bibinfo {pages} {062005} (\bibinfo {year} {2013})},\ \Eprint {https://arxiv.org/abs/1306.5394} {arXiv:1306.5394 [astro-ph.HE]} \BibitemShut {NoStop}%
\bibitem [{\citenamefont {Gair}\ \emph {et~al.}(2014)\citenamefont {Gair}, \citenamefont {Romano}, \citenamefont {Taylor},\ and\ \citenamefont {Mingarelli}}]{Gair:2014rwa}%
  \BibitemOpen
  \bibfield  {author} {\bibinfo {author} {\bibfnamefont {J.}~\bibnamefont {Gair}}, \bibinfo {author} {\bibfnamefont {J.~D.}\ \bibnamefont {Romano}}, \bibinfo {author} {\bibfnamefont {S.}~\bibnamefont {Taylor}},\ and\ \bibinfo {author} {\bibfnamefont {C.~M.~F.}\ \bibnamefont {Mingarelli}},\ }\bibfield  {title} {\bibinfo {title} {{Mapping gravitational-wave backgrounds using methods from CMB analysis: Application to pulsar timing arrays}},\ }\href {https://doi.org/10.1103/PhysRevD.90.082001} {\bibfield  {journal} {\bibinfo  {journal} {Phys. Rev. D}\ }\textbf {\bibinfo {volume} {90}},\ \bibinfo {pages} {082001} (\bibinfo {year} {2014})},\ \Eprint {https://arxiv.org/abs/1406.4664} {arXiv:1406.4664 [gr-qc]} \BibitemShut {NoStop}%
\bibitem [{\citenamefont {Taylor}\ and\ \citenamefont {Gair}(2013)}]{Taylor:2013esa}%
  \BibitemOpen
  \bibfield  {author} {\bibinfo {author} {\bibfnamefont {S.~R.}\ \bibnamefont {Taylor}}\ and\ \bibinfo {author} {\bibfnamefont {J.~R.}\ \bibnamefont {Gair}},\ }\bibfield  {title} {\bibinfo {title} {{Searching For Anisotropic Gravitational-wave Backgrounds Using Pulsar Timing Arrays}},\ }\href {https://doi.org/10.1103/PhysRevD.88.084001} {\bibfield  {journal} {\bibinfo  {journal} {Phys. Rev. D}\ }\textbf {\bibinfo {volume} {88}},\ \bibinfo {pages} {084001} (\bibinfo {year} {2013})},\ \Eprint {https://arxiv.org/abs/1306.5395} {arXiv:1306.5395 [gr-qc]} \BibitemShut {NoStop}%
\bibitem [{\citenamefont {Agazie}\ \emph {et~al.}(2023{\natexlab{b}})\citenamefont {Agazie} \emph {et~al.}}]{NANOGrav:2023tcn}%
  \BibitemOpen
  \bibfield  {author} {\bibinfo {author} {\bibfnamefont {G.}~\bibnamefont {Agazie}} \emph {et~al.} (\bibinfo {collaboration} {NANOGrav}),\ }\bibfield  {title} {\bibinfo {title} {{The NANOGrav 15 yr Data Set: Search for Anisotropy in the Gravitational-wave Background}},\ }\href {https://doi.org/10.3847/2041-8213/acf4fd} {\bibfield  {journal} {\bibinfo  {journal} {Astrophys. J. Lett.}\ }\textbf {\bibinfo {volume} {956}},\ \bibinfo {pages} {L3} (\bibinfo {year} {2023}{\natexlab{b}})},\ \Eprint {https://arxiv.org/abs/2306.16221} {arXiv:2306.16221 [astro-ph.HE]} \BibitemShut {NoStop}%
\bibitem [{\citenamefont {Chu}\ \emph {et~al.}(2021)\citenamefont {Chu}, \citenamefont {Liu},\ and\ \citenamefont {Ng}}]{Chu:2021krj}%
  \BibitemOpen
  \bibfield  {author} {\bibinfo {author} {\bibfnamefont {Y.-K.}\ \bibnamefont {Chu}}, \bibinfo {author} {\bibfnamefont {G.-C.}\ \bibnamefont {Liu}},\ and\ \bibinfo {author} {\bibfnamefont {K.-W.}\ \bibnamefont {Ng}},\ }\bibfield  {title} {\bibinfo {title} {{Observation of a polarized stochastic gravitational-wave background in pulsar-timing-array experiments}},\ }\href {https://doi.org/10.1103/PhysRevD.104.124018} {\bibfield  {journal} {\bibinfo  {journal} {Phys. Rev. D}\ }\textbf {\bibinfo {volume} {104}},\ \bibinfo {pages} {124018} (\bibinfo {year} {2021})},\ \Eprint {https://arxiv.org/abs/2107.00536} {arXiv:2107.00536 [gr-qc]} \BibitemShut {NoStop}%
\bibitem [{\citenamefont {Liu}\ and\ \citenamefont {Ng}(2022)}]{Liu:2022skj}%
  \BibitemOpen
  \bibfield  {author} {\bibinfo {author} {\bibfnamefont {G.-C.}\ \bibnamefont {Liu}}\ and\ \bibinfo {author} {\bibfnamefont {K.-W.}\ \bibnamefont {Ng}},\ }\bibfield  {title} {\bibinfo {title} {{Timing-residual power spectrum of a polarized stochastic gravitational-wave background in pulsar-timing-array observation}},\ }\href {https://doi.org/10.1103/PhysRevD.106.064004} {\bibfield  {journal} {\bibinfo  {journal} {Phys. Rev. D}\ }\textbf {\bibinfo {volume} {106}},\ \bibinfo {pages} {064004} (\bibinfo {year} {2022})},\ \Eprint {https://arxiv.org/abs/2201.06767} {arXiv:2201.06767 [gr-qc]} \BibitemShut {NoStop}%
\bibitem [{\citenamefont {Anil~Kumar}\ \emph {et~al.}(2023)\citenamefont {Anil~Kumar}, \citenamefont {\c{C}al\i{}\c{s}kan}, \citenamefont {Sato-Polito}, \citenamefont {Kamionkowski},\ and\ \citenamefont {Ji}}]{AnilKumar:2023kvt}%
  \BibitemOpen
  \bibfield  {author} {\bibinfo {author} {\bibfnamefont {N.}~\bibnamefont {Anil~Kumar}}, \bibinfo {author} {\bibfnamefont {M.}~\bibnamefont {\c{C}al\i{}\c{s}kan}}, \bibinfo {author} {\bibfnamefont {G.}~\bibnamefont {Sato-Polito}}, \bibinfo {author} {\bibfnamefont {M.}~\bibnamefont {Kamionkowski}},\ and\ \bibinfo {author} {\bibfnamefont {L.}~\bibnamefont {Ji}},\ }\bibfield  {title} {\bibinfo {title} {{Linear polarization of the stochastic gravitational-wave background with pulsar timing arrays}},\ }\href@noop {} {\  (\bibinfo {year} {2023})},\ \Eprint {https://arxiv.org/abs/2312.03056} {arXiv:2312.03056 [astro-ph.CO]} \BibitemShut {NoStop}%
\bibitem [{\citenamefont {Kato}\ and\ \citenamefont {Soda}(2016)}]{Kato:2015bye}%
  \BibitemOpen
  \bibfield  {author} {\bibinfo {author} {\bibfnamefont {R.}~\bibnamefont {Kato}}\ and\ \bibinfo {author} {\bibfnamefont {J.}~\bibnamefont {Soda}},\ }\bibfield  {title} {\bibinfo {title} {{Probing circular polarization in stochastic gravitational wave background with pulsar timing arrays}},\ }\href {https://doi.org/10.1103/PhysRevD.93.062003} {\bibfield  {journal} {\bibinfo  {journal} {Phys. Rev. D}\ }\textbf {\bibinfo {volume} {93}},\ \bibinfo {pages} {062003} (\bibinfo {year} {2016})},\ \Eprint {https://arxiv.org/abs/1512.09139} {arXiv:1512.09139 [gr-qc]} \BibitemShut {NoStop}%
\bibitem [{\citenamefont {Sato-Polito}\ and\ \citenamefont {Kamionkowski}(2022)}]{Sato-Polito:2021efu}%
  \BibitemOpen
  \bibfield  {author} {\bibinfo {author} {\bibfnamefont {G.}~\bibnamefont {Sato-Polito}}\ and\ \bibinfo {author} {\bibfnamefont {M.}~\bibnamefont {Kamionkowski}},\ }\bibfield  {title} {\bibinfo {title} {{Pulsar-timing measurement of the circular polarization of the stochastic gravitational-wave background}},\ }\href {https://doi.org/10.1103/PhysRevD.106.023004} {\bibfield  {journal} {\bibinfo  {journal} {Phys. Rev. D}\ }\textbf {\bibinfo {volume} {106}},\ \bibinfo {pages} {023004} (\bibinfo {year} {2022})},\ \Eprint {https://arxiv.org/abs/2111.05867} {arXiv:2111.05867 [astro-ph.CO]} \BibitemShut {NoStop}%
\bibitem [{\citenamefont {Conneely}\ \emph {et~al.}(2019)\citenamefont {Conneely}, \citenamefont {Jaffe},\ and\ \citenamefont {Mingarelli}}]{Conneely:2018wis}%
  \BibitemOpen
  \bibfield  {author} {\bibinfo {author} {\bibfnamefont {C.}~\bibnamefont {Conneely}}, \bibinfo {author} {\bibfnamefont {A.~H.}\ \bibnamefont {Jaffe}},\ and\ \bibinfo {author} {\bibfnamefont {C.~M.~F.}\ \bibnamefont {Mingarelli}},\ }\bibfield  {title} {\bibinfo {title} {{On the Amplitude and Stokes Parameters of a Stochastic Gravitational-Wave Background}},\ }\href {https://doi.org/10.1093/mnras/stz1022} {\bibfield  {journal} {\bibinfo  {journal} {Mon. Not. Roy. Astron. Soc.}\ }\textbf {\bibinfo {volume} {487}},\ \bibinfo {pages} {562} (\bibinfo {year} {2019})},\ \Eprint {https://arxiv.org/abs/1808.05920} {arXiv:1808.05920 [astro-ph.CO]} \BibitemShut {NoStop}%
\bibitem [{\citenamefont {Ali-Ha\"\i{}moud}\ \emph {et~al.}(2021)\citenamefont {Ali-Ha\"\i{}moud}, \citenamefont {Smith},\ and\ \citenamefont {Mingarelli}}]{Ali-Haimoud:2020iyz}%
  \BibitemOpen
  \bibfield  {author} {\bibinfo {author} {\bibfnamefont {Y.}~\bibnamefont {Ali-Ha\"\i{}moud}}, \bibinfo {author} {\bibfnamefont {T.~L.}\ \bibnamefont {Smith}},\ and\ \bibinfo {author} {\bibfnamefont {C.~M.~F.}\ \bibnamefont {Mingarelli}},\ }\bibfield  {title} {\bibinfo {title} {{Insights into searches for anisotropies in the nanohertz gravitational-wave background}},\ }\href {https://doi.org/10.1103/PhysRevD.103.042009} {\bibfield  {journal} {\bibinfo  {journal} {Phys. Rev. D}\ }\textbf {\bibinfo {volume} {103}},\ \bibinfo {pages} {042009} (\bibinfo {year} {2021})},\ \Eprint {https://arxiv.org/abs/2010.13958} {arXiv:2010.13958 [gr-qc]} \BibitemShut {NoStop}%
\bibitem [{\citenamefont {Ali-Ha\"\i{}moud}\ \emph {et~al.}(2020)\citenamefont {Ali-Ha\"\i{}moud}, \citenamefont {Smith},\ and\ \citenamefont {Mingarelli}}]{Ali-Haimoud:2020ozu}%
  \BibitemOpen
  \bibfield  {author} {\bibinfo {author} {\bibfnamefont {Y.}~\bibnamefont {Ali-Ha\"\i{}moud}}, \bibinfo {author} {\bibfnamefont {T.~L.}\ \bibnamefont {Smith}},\ and\ \bibinfo {author} {\bibfnamefont {C.~M.~F.}\ \bibnamefont {Mingarelli}},\ }\bibfield  {title} {\bibinfo {title} {{Fisher formalism for anisotropic gravitational-wave background searches with pulsar timing arrays}},\ }\href {https://doi.org/10.1103/PhysRevD.102.122005} {\bibfield  {journal} {\bibinfo  {journal} {Phys. Rev. D}\ }\textbf {\bibinfo {volume} {102}},\ \bibinfo {pages} {122005} (\bibinfo {year} {2020})},\ \Eprint {https://arxiv.org/abs/2006.14570} {arXiv:2006.14570 [gr-qc]} \BibitemShut {NoStop}%
\bibitem [{\citenamefont {Hajian}\ and\ \citenamefont {Souradeep}(2003)}]{Hajian:2003qq}%
  \BibitemOpen
  \bibfield  {author} {\bibinfo {author} {\bibfnamefont {A.}~\bibnamefont {Hajian}}\ and\ \bibinfo {author} {\bibfnamefont {T.}~\bibnamefont {Souradeep}},\ }\bibfield  {title} {\bibinfo {title} {{Measuring statistical isotropy of the CMB anisotropy}},\ }\href {https://doi.org/10.1086/379757} {\bibfield  {journal} {\bibinfo  {journal} {Astrophys. J. Lett.}\ }\textbf {\bibinfo {volume} {597}},\ \bibinfo {pages} {L5} (\bibinfo {year} {2003})},\ \Eprint {https://arxiv.org/abs/astro-ph/0308001} {arXiv:astro-ph/0308001} \BibitemShut {NoStop}%
\bibitem [{\citenamefont {{Hajian}}\ and\ \citenamefont {{Souradeep}}(2004)}]{Hajian:2005jh}%
  \BibitemOpen
  \bibfield  {author} {\bibinfo {author} {\bibfnamefont {A.}~\bibnamefont {{Hajian}}}\ and\ \bibinfo {author} {\bibfnamefont {T.}~\bibnamefont {{Souradeep}}},\ }\bibfield  {title} {\bibinfo {title} {{The Cosmic Microwave Background Bipolar Power Spectrum: Basic Formalism and Applications}},\ }\href {https://doi.org/10.48550/arXiv.astro-ph/0501001} {\bibfield  {journal} {\bibinfo  {journal} {arXiv e-prints}\ ,\ \bibinfo {eid} {astro-ph/0501001}} (\bibinfo {year} {2004})},\ \Eprint {https://arxiv.org/abs/astro-ph/0501001} {arXiv:astro-ph/0501001 [astro-ph]} \BibitemShut {NoStop}%
\bibitem [{\citenamefont {Joshi}\ \emph {et~al.}(2010)\citenamefont {Joshi}, \citenamefont {Jhingan}, \citenamefont {Souradeep},\ and\ \citenamefont {Hajian}}]{Joshi:2009mj}%
  \BibitemOpen
  \bibfield  {author} {\bibinfo {author} {\bibfnamefont {N.}~\bibnamefont {Joshi}}, \bibinfo {author} {\bibfnamefont {S.}~\bibnamefont {Jhingan}}, \bibinfo {author} {\bibfnamefont {T.}~\bibnamefont {Souradeep}},\ and\ \bibinfo {author} {\bibfnamefont {A.}~\bibnamefont {Hajian}},\ }\bibfield  {title} {\bibinfo {title} {{Bipolar Harmonic encoding of CMB correlation patterns}},\ }\href {https://doi.org/10.1103/PhysRevD.81.083012} {\bibfield  {journal} {\bibinfo  {journal} {Phys. Rev. D}\ }\textbf {\bibinfo {volume} {81}},\ \bibinfo {pages} {083012} (\bibinfo {year} {2010})},\ \Eprint {https://arxiv.org/abs/0912.3217} {arXiv:0912.3217 [astro-ph.CO]} \BibitemShut {NoStop}%
\bibitem [{\citenamefont {Book}\ \emph {et~al.}(2012)\citenamefont {Book}, \citenamefont {Kamionkowski},\ and\ \citenamefont {Souradeep}}]{Book:2011na}%
  \BibitemOpen
  \bibfield  {author} {\bibinfo {author} {\bibfnamefont {L.~G.}\ \bibnamefont {Book}}, \bibinfo {author} {\bibfnamefont {M.}~\bibnamefont {Kamionkowski}},\ and\ \bibinfo {author} {\bibfnamefont {T.}~\bibnamefont {Souradeep}},\ }\bibfield  {title} {\bibinfo {title} {{Odd-Parity Bipolar Spherical Harmonics}},\ }\href {https://doi.org/10.1103/PhysRevD.85.023010} {\bibfield  {journal} {\bibinfo  {journal} {Phys. Rev. D}\ }\textbf {\bibinfo {volume} {85}},\ \bibinfo {pages} {023010} (\bibinfo {year} {2012})},\ \Eprint {https://arxiv.org/abs/1109.2910} {arXiv:1109.2910 [astro-ph.CO]} \BibitemShut {NoStop}%
\bibitem [{\citenamefont {Belgacem}\ and\ \citenamefont {Kamionkowski}(2020)}]{Belgacem:2020nda}%
  \BibitemOpen
  \bibfield  {author} {\bibinfo {author} {\bibfnamefont {E.}~\bibnamefont {Belgacem}}\ and\ \bibinfo {author} {\bibfnamefont {M.}~\bibnamefont {Kamionkowski}},\ }\bibfield  {title} {\bibinfo {title} {{Chirality of the gravitational-wave background and pulsar-timing arrays}},\ }\href {https://doi.org/10.1103/PhysRevD.102.023004} {\bibfield  {journal} {\bibinfo  {journal} {Phys. Rev. D}\ }\textbf {\bibinfo {volume} {102}},\ \bibinfo {pages} {023004} (\bibinfo {year} {2020})},\ \Eprint {https://arxiv.org/abs/2004.05480} {arXiv:2004.05480 [astro-ph.CO]} \BibitemShut {NoStop}%
\bibitem [{\citenamefont {Roebber}\ and\ \citenamefont {Holder}(2017)}]{Roebber:2016jzl}%
  \BibitemOpen
  \bibfield  {author} {\bibinfo {author} {\bibfnamefont {E.}~\bibnamefont {Roebber}}\ and\ \bibinfo {author} {\bibfnamefont {G.}~\bibnamefont {Holder}},\ }\bibfield  {title} {\bibinfo {title} {{Harmonic space analysis of pulsar timing array redshift maps}},\ }\href {https://doi.org/10.3847/1538-4357/835/1/21} {\bibfield  {journal} {\bibinfo  {journal} {Astrophys. J.}\ }\textbf {\bibinfo {volume} {835}},\ \bibinfo {pages} {21} (\bibinfo {year} {2017})},\ \Eprint {https://arxiv.org/abs/1609.06758} {arXiv:1609.06758 [astro-ph.CO]} \BibitemShut {NoStop}%
\bibitem [{\citenamefont {Hotinli}\ \emph {et~al.}(2019)\citenamefont {Hotinli}, \citenamefont {Kamionkowski},\ and\ \citenamefont {Jaffe}}]{Hotinli:2019tpc}%
  \BibitemOpen
  \bibfield  {author} {\bibinfo {author} {\bibfnamefont {S.~C.}\ \bibnamefont {Hotinli}}, \bibinfo {author} {\bibfnamefont {M.}~\bibnamefont {Kamionkowski}},\ and\ \bibinfo {author} {\bibfnamefont {A.~H.}\ \bibnamefont {Jaffe}},\ }\bibfield  {title} {\bibinfo {title} {{The search for anisotropy in the gravitational-wave background with pulsar-timing arrays}},\ }\href {https://doi.org/10.21105/astro.1904.05348} {\bibfield  {journal} {\bibinfo  {journal} {Open J. Astrophys.}\ }\textbf {\bibinfo {volume} {2}},\ \bibinfo {pages} {8} (\bibinfo {year} {2019})},\ \Eprint {https://arxiv.org/abs/1904.05348} {arXiv:1904.05348 [astro-ph.CO]} \BibitemShut {NoStop}%
\bibitem [{\citenamefont {Arzoumanian}\ \emph {et~al.}(2021)\citenamefont {Arzoumanian} \emph {et~al.}}]{NANOGrav:2021ini}%
  \BibitemOpen
  \bibfield  {author} {\bibinfo {author} {\bibfnamefont {Z.}~\bibnamefont {Arzoumanian}} \emph {et~al.} (\bibinfo {collaboration} {NANOGrav}),\ }\bibfield  {title} {\bibinfo {title} {{The NANOGrav 12.5-year Data Set: Search for Non-Einsteinian Polarization Modes in the Gravitational-wave Background}},\ }\href {https://doi.org/10.3847/2041-8213/ac401c} {\bibfield  {journal} {\bibinfo  {journal} {Astrophys. J. Lett.}\ }\textbf {\bibinfo {volume} {923}},\ \bibinfo {pages} {L22} (\bibinfo {year} {2021})},\ \Eprint {https://arxiv.org/abs/2109.14706} {arXiv:2109.14706 [gr-qc]} \BibitemShut {NoStop}%
\bibitem [{\citenamefont {Bernardo}\ and\ \citenamefont {Ng}(2023{\natexlab{b}})}]{Bernardo:2022rif}%
  \BibitemOpen
  \bibfield  {author} {\bibinfo {author} {\bibfnamefont {R.~C.}\ \bibnamefont {Bernardo}}\ and\ \bibinfo {author} {\bibfnamefont {K.-W.}\ \bibnamefont {Ng}},\ }\bibfield  {title} {\bibinfo {title} {{Stochastic gravitational wave background phenomenology in a pulsar timing array}},\ }\href {https://doi.org/10.1103/PhysRevD.107.044007} {\bibfield  {journal} {\bibinfo  {journal} {Phys. Rev. D}\ }\textbf {\bibinfo {volume} {107}},\ \bibinfo {pages} {044007} (\bibinfo {year} {2023}{\natexlab{b}})},\ \Eprint {https://arxiv.org/abs/2208.12538} {arXiv:2208.12538 [gr-qc]} \BibitemShut {NoStop}%
\bibitem [{\citenamefont {Book}\ and\ \citenamefont {Flanagan}(2011)}]{Book:2010pf}%
  \BibitemOpen
  \bibfield  {author} {\bibinfo {author} {\bibfnamefont {L.~G.}\ \bibnamefont {Book}}\ and\ \bibinfo {author} {\bibfnamefont {E.~E.}\ \bibnamefont {Flanagan}},\ }\bibfield  {title} {\bibinfo {title} {{Astrometric Effects of a Stochastic Gravitational Wave Background}},\ }\href {https://doi.org/10.1103/PhysRevD.83.024024} {\bibfield  {journal} {\bibinfo  {journal} {Phys. Rev. D}\ }\textbf {\bibinfo {volume} {83}},\ \bibinfo {pages} {024024} (\bibinfo {year} {2011})},\ \Eprint {https://arxiv.org/abs/1009.4192} {arXiv:1009.4192 [astro-ph.CO]} \BibitemShut {NoStop}%
\end{thebibliography}%
\end{document}